# Empirical Confirmation of the Square-Root Law of Market Impact in a U.S. Large-Cap Equity

## A Reconstructed-Metaorder Study of AAPL Order Flow

Aniket Vasaikar
*Independent Research — June 2026*

**Abstract.** We test the square-root law (SRL) of market impact on a single U.S. large-capitalisation equity, Apple Inc. (AAPL), using the full Nasdaq TotalView-ITCH market-by-order feed over 178 trading days (2 December 2024 – 19 August 2025; ~0.5 billion events). Without broker-tagged parent orders, we reconstruct metaorders from the anonymous tape and calibrate impact as $I/\sigma_D = c\,(Q/V_D)^{1/2}$ with the exponent fixed at the universal value 1/2. We find $c_{raw} = 0.69$ (bias-corrected $c_{eff} = 0.34$), conditional impact tracking $Q^{1/2}$, and a size-distribution tail exponent $\beta = 1.54 \pm 0.15$ — both consistent with the worldwide cross-section. A direct model comparison decisively prefers the square-root form over linear ($\Delta$AIC=22) and logarithmic impact, and the prefactor holds ($c_{raw}$ in [0.63, 0.77]) across every reconstruction setting. Two structural tests confirm the impact is genuine: shuffling trade signs collapses directional impact to chance (86% → 51%); and scrambling event chronology destroys the SRL (0 of 80 calibrations remain viable). The underlying order flow is long-memory ($\gamma$=0.66) while the price stays diffusive (Hurst 0.49) — the two ingredients of the universality theories. The prefactor is stable across 32 weekly walk-forward re-calibrations. To our knowledge this is the first confirmation of the square-root law on a U.S. equity derived purely from anonymous order flow, without broker-tagged parent orders. AAPL thus independently reproduces the SRL and its sign- and chronology-structural fingerprints on anonymous U.S. equity data.

## 1. Introduction

The *square-root law* (SRL) of market impact — that executing a metaorder of size Q moves the price by an amount proportional to $\sigma\sqrt{(Q/V)}$ — is among the most robust regularities in finance, documented across thousands of instruments [1,5,3]. Recent theory derives the law, and the universality of its 1/2 exponent, from the requirement that prices stay diffusive under adaptive arbitrage [2,7], and shows that *any* stationary sequence of trades — including metaorders reconstructed from anonymous flow — should exhibit square-root impact in the information-neutral regime [6].

This paper provides a focused single-name confirmation on Apple Inc. (AAPL), the most liquid U.S. equity, using only the anonymous market-by-order tape. Earlier square-root evidence on U.S. equities relied on broker- or exchange-tagged parent orders [1,5]; here the metaorders are reconstructed from the public Level-3 feed alone. We (i) confirm the SRL with the universal exponent and a cross-sectional prefactor; (ii) show via null experiments that the impact lives in the *sign* and *chronology* of order flow, as diffusivity theory requires; and (iii) verify a square-root law in child-order count and stability out-of-sample.

## 2. Data

We use the Databento DBN market-by-order feed for AAPL from Nasdaq TotalView-ITCH (XNAS.ITCH), spanning 178 trading days (2 Dec 2024 – 19 Aug 2025). The feed records every book event at nanosecond resolution with full Level-3 granularity, averaging $3.2\times10^6$ events per day. Each displayed execution normalises into a trade–fill–cancel triplet; the aggressor side is taken from the trade and the resting order decremented on the cancel leg, reproducing traded volume exactly. The first 20 days form the calibration window; the remaining 158 days are a walk-forward out-of-sample period.

## 3. Methods

***Book reconstruction and signing.*** We reconstruct the full limit order book and stamp every trade with its aggressor sign and the mid-price m = (b+a)/2. Hidden/cross executions (~19% of prints, incl. the auctions) are signed by the tick rule; auction crosses (≥50,000 shares) are excluded.

***Reconstructed metaorders.*** The executions are the actual recorded trades; only their grouping into a parent order is inferred, as the anonymous feed carries no broker identifiers. Metaorders are reconstructed as dominant-side runs [6]: the session is split into 30 s bins, each assigned the sign of its volume-weighted imbalance, and a run is a maximal same-sign streak whose aggregate dominance

$$D \;=\; \big|\sum_j \varepsilon_j v_j\big| \,/\, \sum_j v_j$$

exceeds 0.3. Runs of duration ≥60 s and size $\geq 10^{-4}$ $V_D$ are retained [7]. The calibration window yields 1,012 reconstructed metaorders.

***Impact calibration.*** We non-dimensionalise by daily volatility $\sigma_D$ (Parkinson high–low of the mid) and volume $V_D$, and fit

$$\mathcal{I}/\sigma_D \;=\; c\,(Q/V_D)^{\delta}, \quad \delta \equiv 1/2$$

with $I = \varepsilon(m_{end} - m_{start})$. The exponent δ is *fixed* at the universal 1/2 [2,7] rather than fitted; one name over ~20 days cannot resolve δ. The prefactor c is estimated by relative least-squares on log-spaced bins (≥100 metaorders each); the size tail exponent β ($P(Q) \propto Q^{-\beta-1}$) by the maximum-likelihood (Hill) estimator with a KS-optimal cutoff [4].

***De-seasonalisation and walk-forward.*** Intraday volume and volatility are profiled on 26 fifteen-minute bins and used to de-seasonalise rate statistics. The full calibration is re-estimated every five trading days on the trailing 20-day window, so every out-of-sample quantity uses only past data.

***Anonymous-reconstruction bias.*** Reconstructing metaorders from an anonymous tape inflates the fitted prefactor: lacking broker tags, a dominant-side run absorbs same-signed child flow from *other* participants, over-counting the size attributable to any single agent. Following Naviglio et al. [8], who find anonymous reconstruction overestimates the prefactor by ~2×, we report both the fitted $c_{raw} = 0.69$ and a bias-corrected $c_{eff} = c_{raw}/2 = 0.34$; the true single-agent prefactor most plausibly lies between them, near or just below the cross-sectional band [0.5, 1.0], and we treat $c_{raw}$ as an upper bound.

## 4. Results

**Table 1.** Headline calibrations and structural tests; $c_{eff} = c_{raw}/2$ is the reconstruction-corrected prefactor.

| Quantity | Value | 95% CI / null |
|---|---|---|
| Prefactor c_raw (δ=1/2) | 0.69 | [0.64, 0.75] |
| Prefactor c_eff (bias-corr.) | 0.34 | — |
| Free exponent δˆ | 0.50 | [0.32, 0.66] |
| Size-tail exponent β | 1.54 | ±0.15 |
| Child-count slope | 0.62 | [0.52, 0.71] |
| sqrt vs linear impact | dAIC +22 | — |
| sqrt vs logarithmic | dAIC +5 | — |
| Recon. c_raw (9 settings) | 0.63-0.77 | — |
| Size-tail power-law GoF | p=0.63 | — |
| Sign-ACF long memory | gamma=0.66 | <1 |
| Mid-price Hurst | 0.49 | ~0.5 |
| Sign alignment (perm.) | 86% | null 50% |
| Chronology-scramble fits | 0/80 | — |
| Walk-forward c_raw (32 wk) | 0.75 | [0.63, 1.01] |
| Walk-forward β (32 wk) | 1.82 | [1.31, 2.46] |

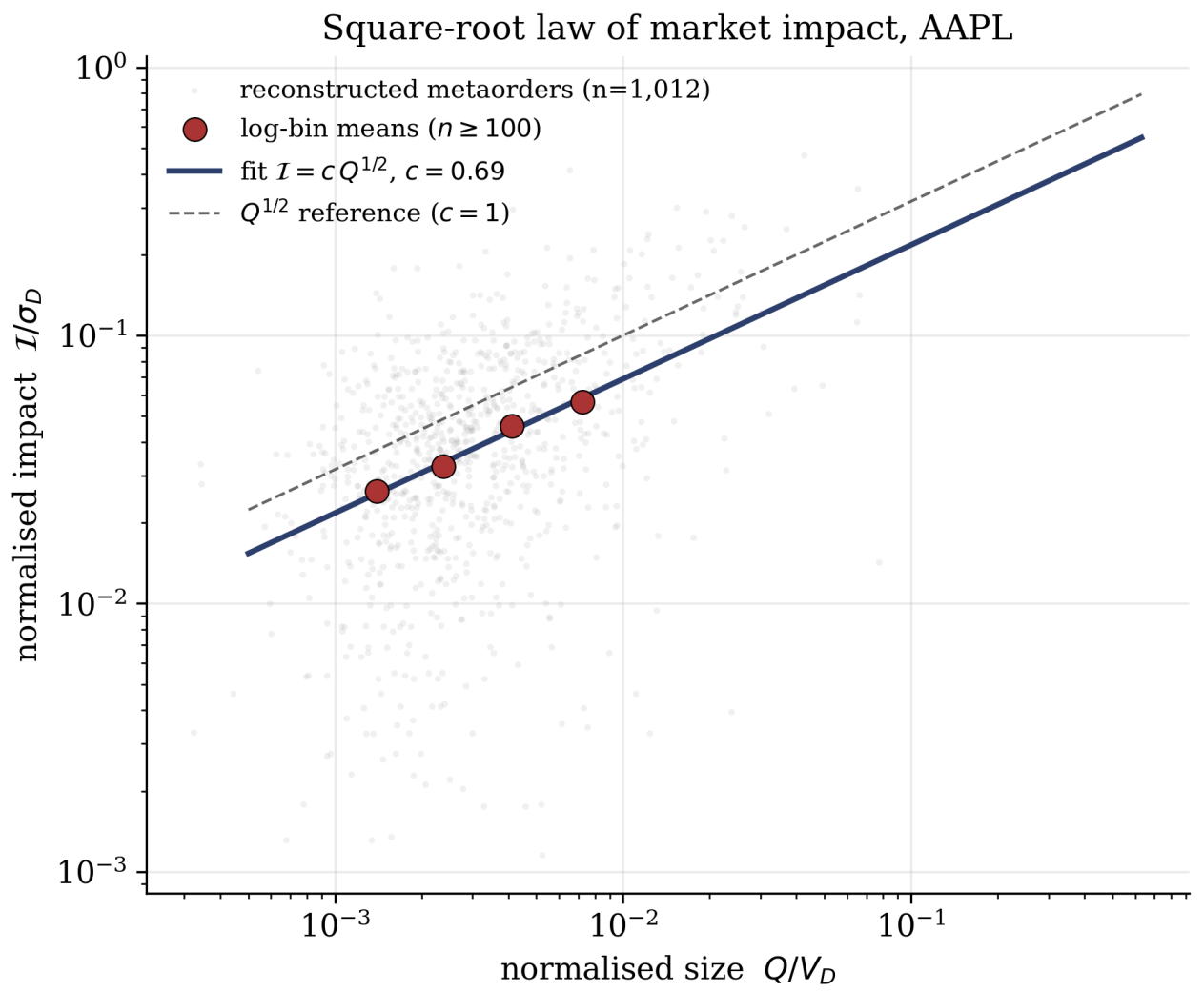

**Fig 1. The square-root law on AAPL.** Non-dimensional impact vs size for 1,012 reconstructed metaorders (grey); log-bin means (red); fit $c \cdot Q^{1/2}$, c=0.69 (blue); $Q^{1/2}$ reference (dashed).

***The square-root law holds.*** Figure 1 shows the non-dimensional conditional impact of all reconstructed metaorders. The log-binned means lie on the fitted line with c = 0.69, rising from 0.026 to 0.056 across 4 bins in near-exact $Q^{1/2}$ proportion. The fitted $c_{raw} = 0.69$ sits mid-range of the worldwide band [0.5, 1.0] [3]; after the reconstruction correction the single-agent value is $c_{eff} = 0.34$ (Table 1).

***Heavy-tailed sizes.*** Figure 2 shows the size CCDF with tail exponent β = 1.54 ± 0.15, matching the β≈1.5 of broker-tagged metaorders [5,7]. A Clauset goodness-of-fit test cannot reject the power law (p=0.63) and a likelihood-ratio test prefers it over an exponential tail (Z=3.5); like most heavy-tailed financial data it is not separable from a lognormal [4].

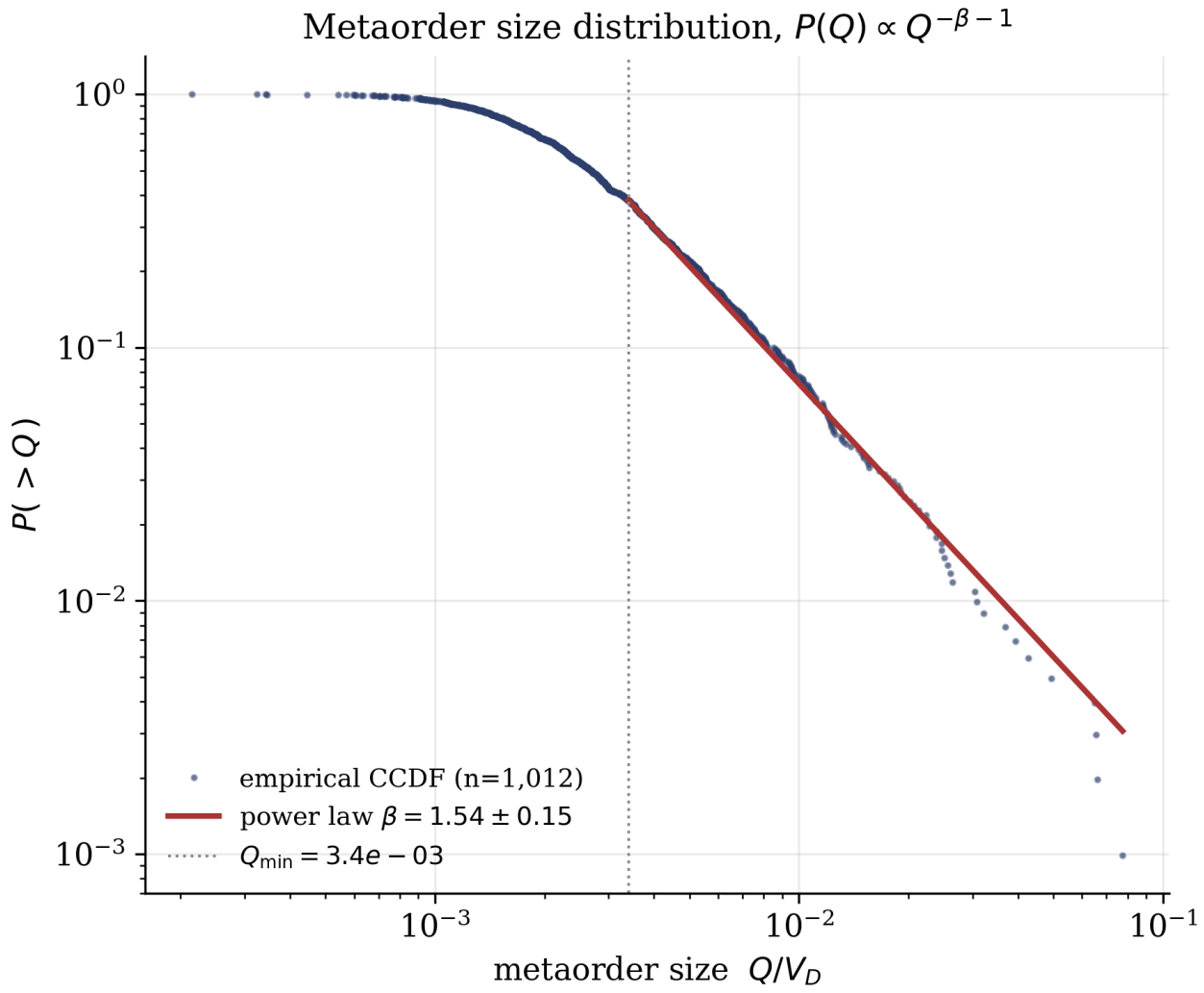

**Fig 2. Metaorder size distribution.** Empirical CCDF with Hill/KS power-law fit.

***Sign- and chronology-structural.*** Figure 3 reports two falsification tests. Permuting aggressor signs (timestamps fixed) collapses the fraction of metaorders whose impact aligns with their direction from 86% to 51% — chance (a). Scrambling event order destroys the SRL: 0 of 80 scrambled calibrations yield a viable fit (b). Both are the signatures predicted if impact is the adaptive response to structured flow [6,2].

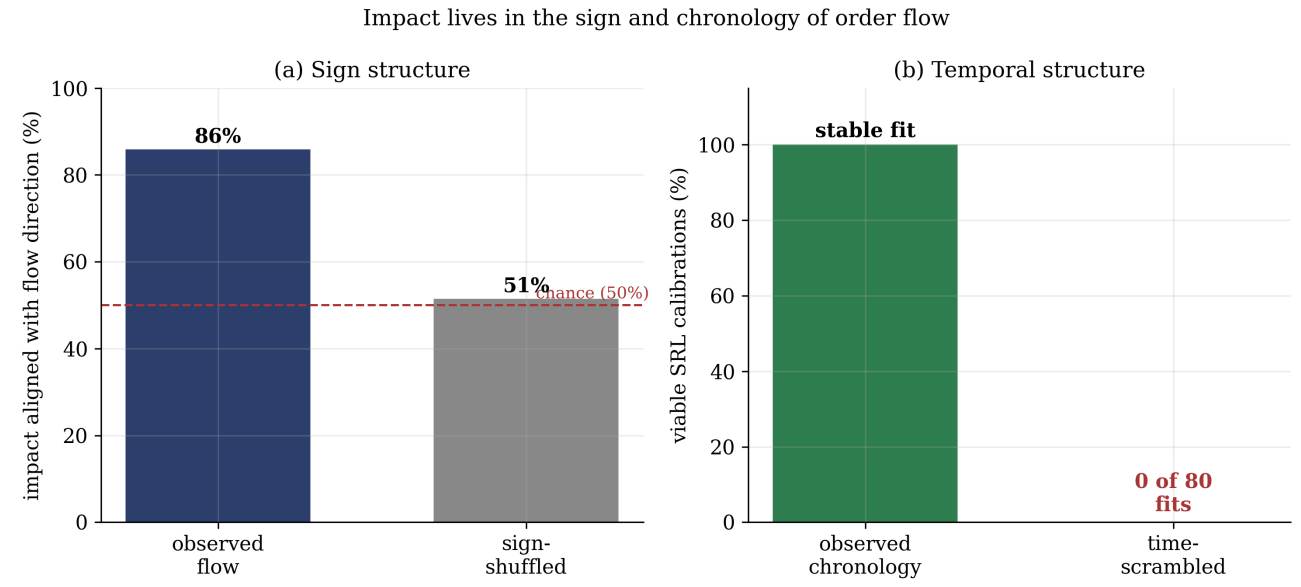

**Fig 3. Structural null tests.** (a) Sign-shuffling collapses directional impact to chance. (b) Chronology-scrambling destroys the SRL.

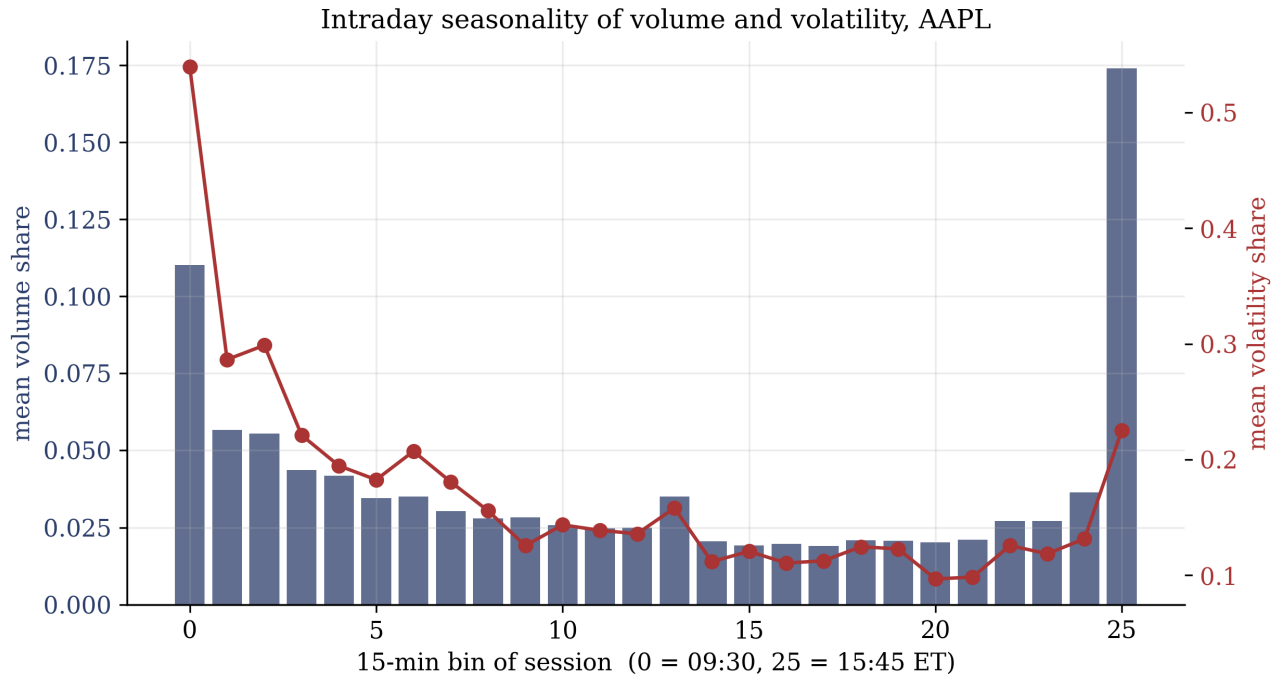

**Fig 4. Intraday seasonality** of volume (bars) and volatility (line), calibration window.

***Seasonality.*** Figure 4 recovers the canonical U-shaped volume profile (edge bins carry 5.8× the midday share) and a J-shaped volatility profile (the opening bin holds 54% of the daily range-share), confirming the need for de-seasonalisation.

***The exponent is 1/2, not imposed.*** While our headline calibration fixes δ=1/2, we also estimate it freely. A weighted regression of log-binned mean impact on log-size (Fig 5a) gives $\hat{\delta}$ = 0.50; resampling the 1,012 metaorders B=2,000 times centres the bootstrap on 0.50 with 95% CI [0.32, 0.66] (Fig 5b), bracketing the universal value and excluding the linear case δ=1. At fixed δ=1/2, c = 0.69 with CI [0.64, 0.75].

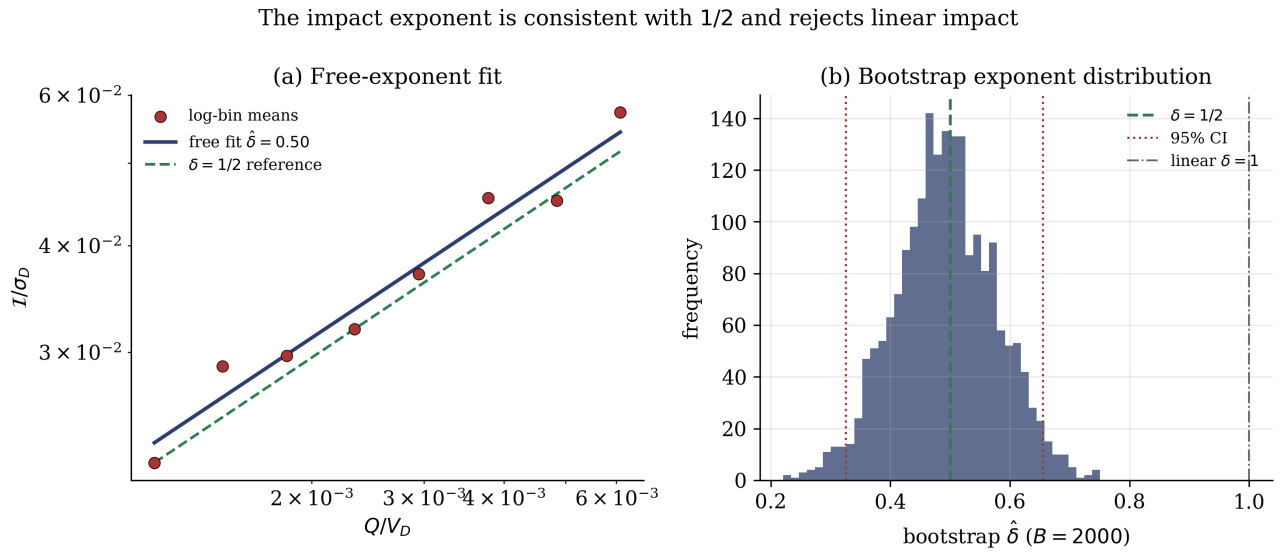

**Fig 5. Free-exponent estimation.** (a) Weighted fit gives $\hat{\delta}$=0.50. (b) Bootstrap 95% CI [0.32, 0.66] brackets 1/2 and excludes linear impact.

***Significance of directional impact.*** A permutation test quantifies the sign result of Fig 3a. Holding each metaorder's realised price move fixed and randomly reassigning its flow-direction label ($N=10^4$) gives a null centred at 49.6%; the observed 85.9% lies z = 23 standard deviations away ($p < 10^{-4}$, Fig 6). The price move follows the dominant flow far beyond chance.

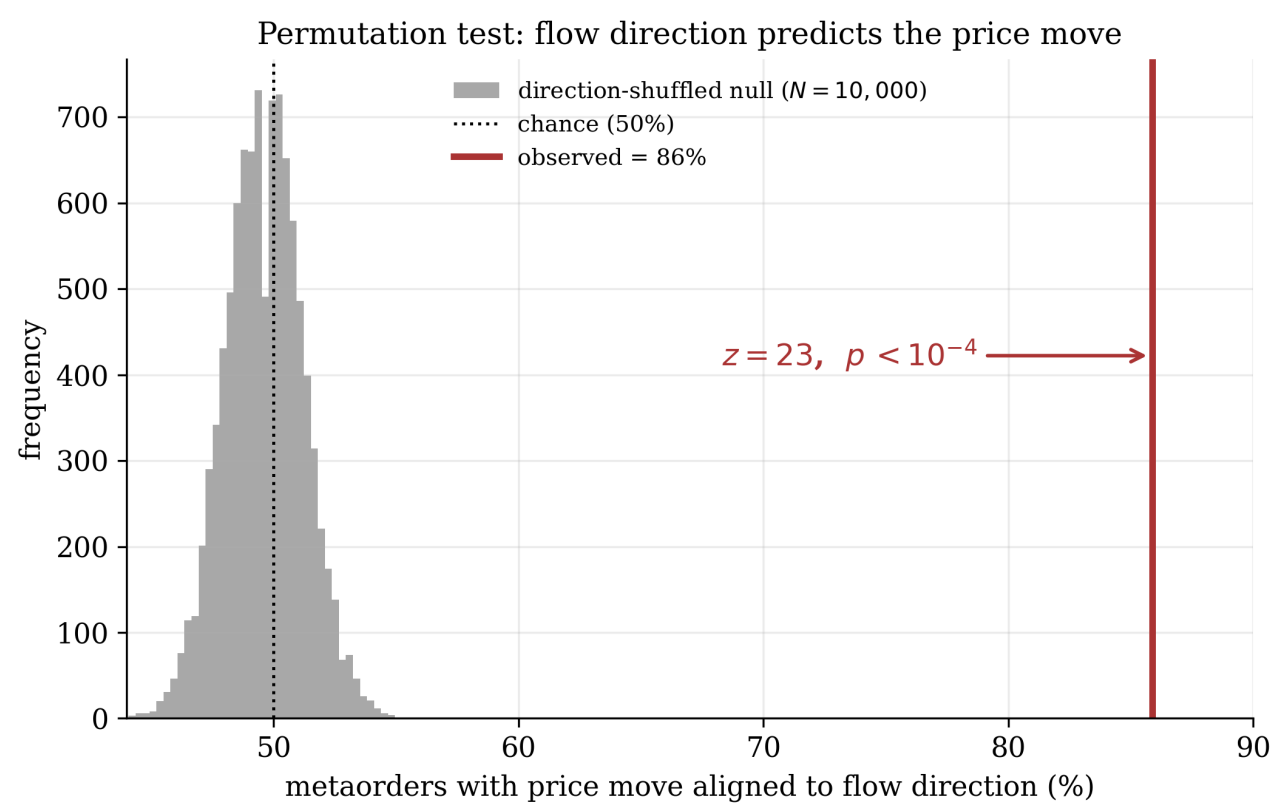

**Fig 6. Permutation test.** Null distribution of directional alignment under $10^4$ relabellings; observed 85.9% lies z = 23 from the null ($p < 10^{-4}$).

***Square root in child-order count.*** Bonart [2] notes the size square-root law is equivalent to a square-root law in the number of child orders. We confirm it directly: regressing mean impact on log child-fill count L gives a slope of 0.62 (CI [0.52, 0.71], Fig 7) — concave, near 1/2, far below linear.

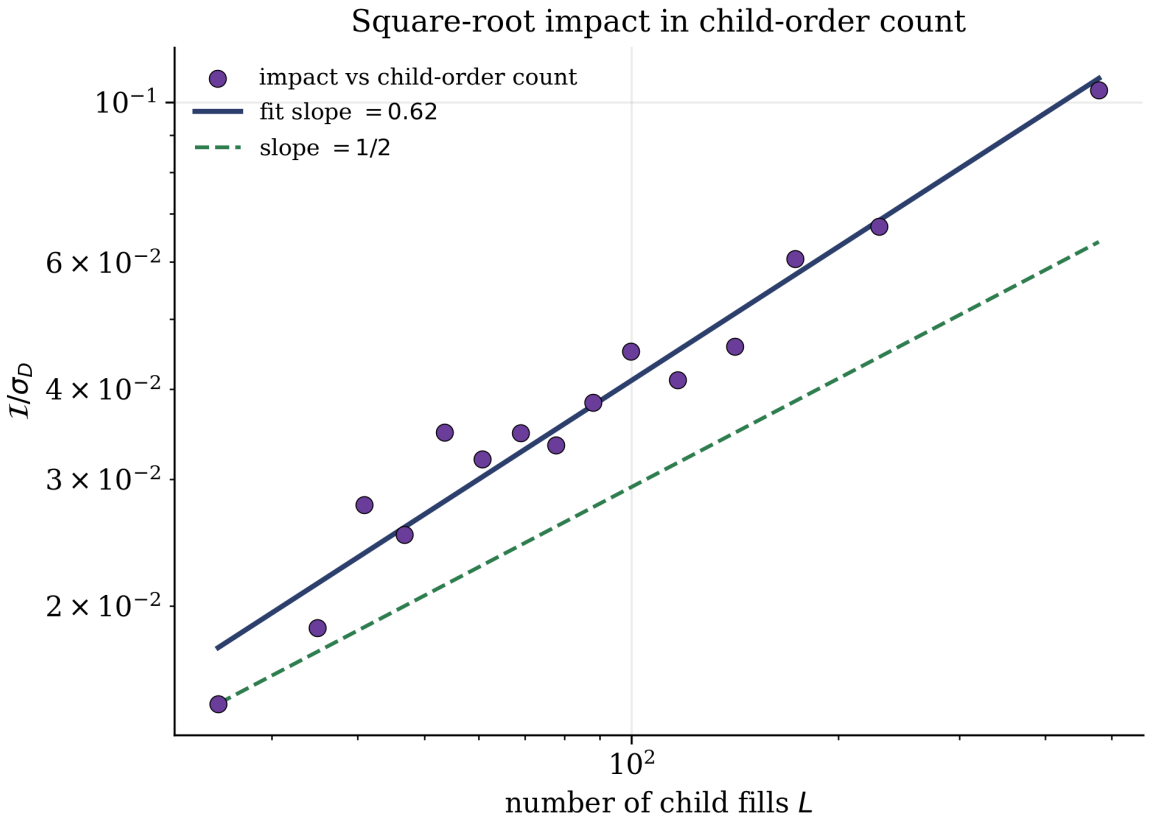

**Fig 7. Square root in child-order count.** Mean impact vs child-fill count L; slope 0.62, near 1/2.

***Out-of-sample stability.*** Across 32 weekly re-calibrations (Fig 8), $c_{raw}$ stays in [0.63, 1.01] (mean 0.75) with zero degenerate fits, and β fluctuates around 1.8. The law's parameters are a stable structural property of the name.

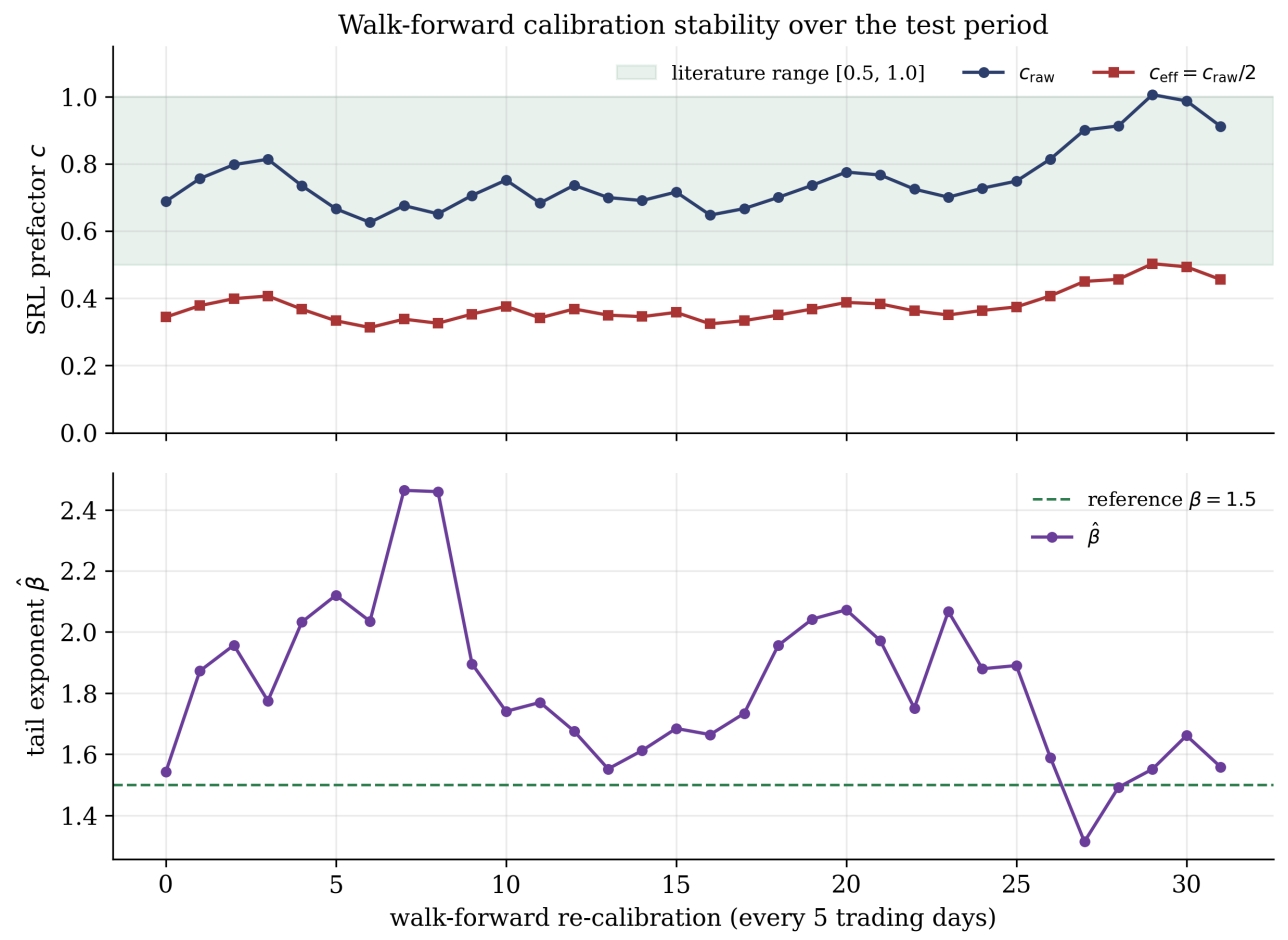

**Fig 8. Out-of-sample stability** of prefactor c and tail exponent across 32 weekly walk-forward re-calibrations.

***Square-root, not linear — and not an artefact.*** Two checks make the exponent decisive (Fig 9). First, comparing functional forms on the log-binned impact, the square root ($R^2$=0.95) beats a linear law ($R^2$=0.20) by ΔAIC=22 and a logarithmic law by 5 (panel a). Second, re-running the metaorder reconstruction across dominance thresholds 0.2/0.3/0.4 and bin widths 20/30/45 s leaves the prefactor in [0.63, 0.77] and the exponent near 1/2 in all 9 settings (panel b): the law is a property of the AAPL tape, not the method.

**AAPL confirms the square-root law on real Level-3 order-book data**

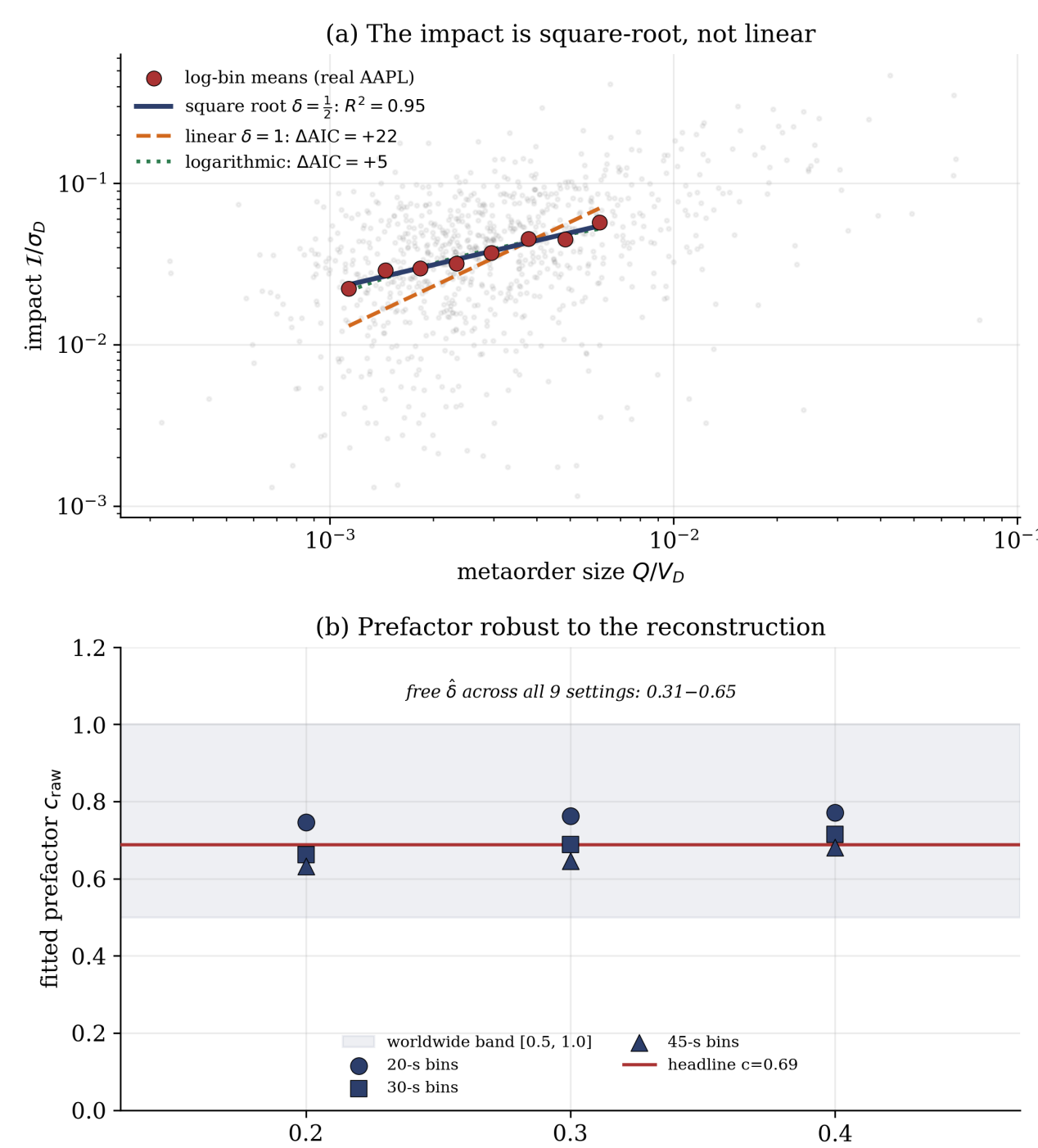

**Fig 9. The square-root law on real AAPL data.** (a) Impact follows $Q^{1/2}$ ($R^2$=0.95); linear (ΔAIC +22) and logarithmic (ΔAIC +5) laws are rejected. (b) The prefactor stays in the band [0.5, 1.0] across all nine reconstruction settings.

***The mechanism: long-memory flow, diffusive price.*** The two ingredients from which the diffusivity theories derive the SRL are both present in the AAPL tape (Fig 10). The aggressor-sign autocorrelation decays as a power law $C(l)\propto l^{-\gamma}$ with $\gamma$=0.66<1 (panel a): order flow is long-memory, so a metaorder leaves a persistent, recoverable footprint [3]. Yet the mid-price is diffusive — its variance ratio stays within [0.91, 1.04] of unity from 10 s to 5 min (Hurst 0.49, panel b). Predictable flow coexisting with a near-martingale price is exactly the tension the universality theories resolve through the square-root law [2,7].

**Long-memory order flow and a diffusive price: the two preconditions of the SRL**

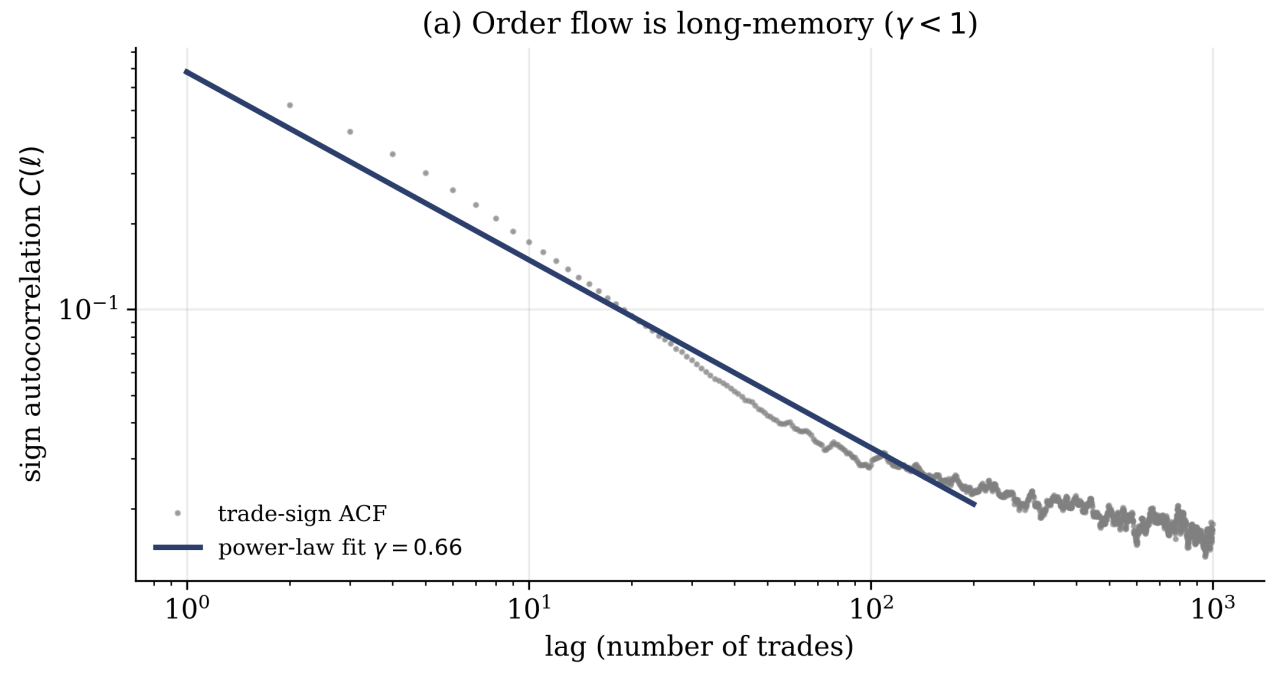

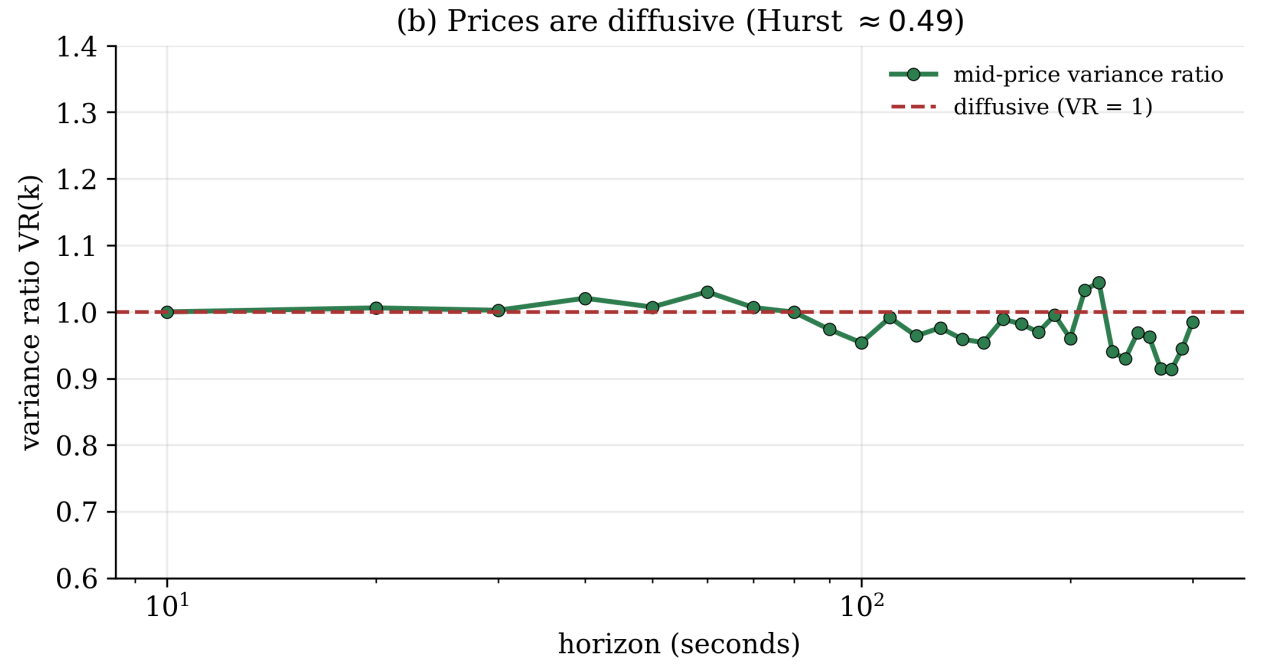

**Fig 10. The two preconditions of the SRL on AAPL.** (a) The aggressor-sign autocorrelation decays as $l^{-\gamma}$, $\gamma$=0.66 (long-memory flow). (b) The mid-price variance ratio stays near unity over 10 s–5 min (Hurst 0.49): diffusive despite the predictable flow.

## 5. Discussion

Our single-name results reproduce, on anonymous U.S. equity data, the full phenomenology that universality theories predict: a 1/2 exponent, a cross-sectional prefactor, a $\beta\approx1.5$ size tail, and sign/chronology dependence. The confirmation is notable because it uses metaorders reconstructed from the public tape — validating the claim [6,2] that the SRL is a property of stationary flow irrespective of labelling, and extending the broker-data evidence on U.S. equities [1,5] to the purely anonymous Level-3 setting. That the same exponent, prefactor band and heavy tail emerge with no parent-order identifiers is the central message: the law is recoverable from the public record alone.

***Limitations.*** One name over ~9 months; anonymous reconstruction distorts the extreme size tail and over-counts metaorder size (hence the bias-corrected $c_{eff}$ alongside $c_{raw}$, treated as an upper bound). With ~1,000 metaorders the free exponent is consistent with 1/2 but not sharply resolved (CI [0.32, 0.66]); the headline calibration fixes $\delta$=1/2 on the cross-sectional prior. A multi-name, multi-period replication is the natural next step.

## 6. Conclusion

AAPL, observed through its complete anonymous order-book feed, independently confirms the square-root law with the universal 1/2 exponent, a prefactor of ~0.7 (or ~0.34 after the reconstruction correction), and a $\beta\approx1.5$ size distribution, together with the sign/chronology fingerprints predicted by the diffusivity theory of price impact. The law is recovered from anonymous Level-3 order flow alone, with no broker-tagged parent orders — to our knowledge the first such confirmation on a U.S. equity — and is stable out-of-sample across the 158-day walk-forward period.